\documentclass[aps,reprint]{revtex4-1}
\usepackage{balance}  % to better equalize the last page
\usepackage{graphicx}
\usepackage{amsmath}
\usepackage{amssymb}
\usepackage{subfigure}
\usepackage{color}

\newcommand{\remove}[1]{}

\begin{document}

%\title{A Meme is not a Virus: Cognitive Heuristics Constrain Information Spread Online}
%\title{A Meme is not a Virus: Cognitive Heuristics Constrain Social Contagion}
\title{Information is not a Virus, and Other Consequences of Human Cognitive Limits}

\author{Kristina Lerman}
\affiliation{Information Sciences Institute, University of Southern California, Marina del Rey, CA 90292}

%\abstract{
\begin{abstract}
The many decisions people make about what to pay attention to online shape
%the collective behavior observed on social media.
the spread of information in online social networks.
Due to the constraints of available time and cognitive resources, the ease of discovery strongly impacts how people allocate their attention to social media content.
As a consequence, the position of information in an individual's social feed, as well as explicit social signals about its popularity, determine whether it will be seen,
%while explicit social signal about its popularity increases
and the likelihood that it will be shared with followers.
Accounting for these cognitive limits simplifies mechanics of information diffusion in online social networks and explains puzzling empirical observations: (\emph{i}) information generally fails to spread in social media and (\emph{ii}) highly connected people are less likely to re-share information.
Studies of information diffusion on different social media platforms reviewed here suggest that the interplay between human cognitive limits and network structure
%can dramatically impact the global dynamics of social contagion in networks.
differentiates the spread of information from other social contagions, such as the spread of a virus through a population.
%}

\end{abstract}

%\keyword{social contagion; information diffusion; social media}
\maketitle
\let\clearpage\relax

\section{Introduction}
The spread of information in online social networks is often likened to the spread of a contagious disease. According to this analogy, % metaphor?
information---whether a trending topic, a news story, a song, or a video---behaves much like a virus, ``infecting'' individuals, who then ``expose'' their naive followers by mentioning the topic, sharing the video, or recommending the news story, etc. These followers may, in turn, become ``infected'' by sharing the information, ``exposing'' their own followers, and so on.
%become informed, for example, by mentioning a topic or sharing a video, and ``expose'' their naive followers to the meme. The followers may, in turn, become ``infected'' and ``expose'' their own followers, and so on.
If each person ``infects'' at least one other person, information will keep spreading on the network, resulting in a ``viral'' outbreak, similar to how a spreading virus can create an epidemic that sickens a large portion of the population. The analogy between the spread of a disease and information is the basis of computational methods that attempt to amplify the spread of information in networks by identifying influential ``superspreaders''~\cite{Kempe03,Leskovec06,Leskovec07,Gruhl2009,socdynamics}, and those that make inferences about the network from observations of how information spreads on it~\cite{newman02,Anagnostopoulos08,Rodriguez2010inferring,Rodriguez2014uncovering,Myers12kdd}.

%\begin{figure}[ht]
%\begin{center}
%\subfigure[]{\label{subfig:icm}\includegraphics[width=0.4\textwidth]{figures/sim_ind_cascade2}}
%\subfigure[]{\label{subfig:fsm}\includegraphics[width=0.4\textwidth]{figures/saturated_cascade_model}}
%\end{center}
% \caption{\emph{Size of simulated outbreaks as a function of transmissibility.} (a) Contagions are simulated using the independent cascade model (ICM) on the Digg follower graph (red dots) and a random graph with the same degree distribution (black x). The golden line gives theoretically predicted outbreak sizes. (b) Simulations of social contagion on the Digg follower graph while suppressing response to multiple exposures, as suggested by Fig.~\protect\ref{subfig:digg-response} (black crosses). Actual outbreaks are shown as red dots, while theoretically predicted (gold) line is the same as in (a). Suppressed response to repeated exposures vastly decreases the size of outbreaks. \label{fig:icm} }
% \end{figure}

\begin{figure}[ht]
\begin{center}
\includegraphics[width=0.4\textwidth]{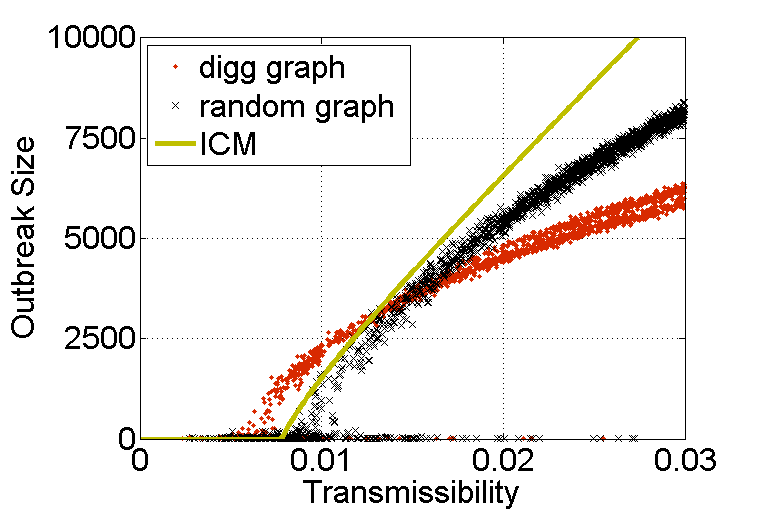}
\end{center}
 \caption{\emph{Size of simulated outbreaks on a real-world and random graphs as a function of transmissibility.} Contagions are simulated using the independent cascade model (ICM) on the follower graph of the Digg social news platform (red dots) and a random graph with the same degree distribution (black crosses). The golden line gives theoretically predicted outbreak sizes.
 %(b) Simulations of social contagion on the Digg follower graph while suppressing response to multiple exposures, as suggested by Fig.~\protect\ref{subfig:digg-response} (black crosses). Actual outbreaks are shown as red dots, while theoretically predicted (gold) line is the same as in (a). Suppressed response to repeated exposures vastly decreases the size of outbreaks.
 \label{fig:icm} }
 \end{figure}

One of the simplest and most widely used models of the spread of epidemics  in networks is the Independent Cascade Model (ICM)~\cite{Hethcote00,Goldenberg01,newman02,Kempe03,Gruhl04}. It describes a  process wherein each exposure of a healthy but susceptible individual to a disease by an infected friend results in an independent chance of disease transmission: the more infected friends an individual has, the more likely he or she is to become infected. The model predicts the size of an outbreak (number of individuals infected) in any network for a given value of disease transmissibility (i.e., how easily the disease is transmitted upon exposure). Figure~\ref{fig:icm} shows the size of outbreaks simulated using the ICM  on a social media follower graph (red dots)~\cite{Versteeg11icwsm}. The black symbols give the size of simulated outbreaks on a randomly-generated graph with the same degree distribution as the follower graph. The simulated outbreaks are close in size to the theoretically predicted values~\cite{moreno}, given by the golden line in Fig.~\ref{fig:icm}. There exists a critical value of transmissibility---the epidemic threshold~\cite{Wang03}---below which the contagion dies out, but above which it spreads to a finite portion of the network.
The epidemic threshold depends only on structural properties of the network, and not the details of the disease or its transmissibility~\cite{Chakrabarti08}: specifically, the epidemic threshold is given by the inverse of the largest eigenvalue of the adjacency matrix representing the network~\cite{Wang03,Ghosh11physrev}.
Note that even above the epidemic threshold, contagions starting in isolated corners of the network may die out. However, in general, the higher the transmissibility, the farther the contagion spreads, reaching a non-negligible fraction of the network above the epidemic threshold, for example, 10\%, 20\%, etc. of the network.

How well does the  \emph{social contagion} analogy hold for social media?
In this review of empirical studies of information diffusion in social media, I first present evidence that information fails to spread widely in online social networks. The vast majority of outbreaks are very small (see Fig.~\ref{fig:outbreak}),  in stark contrast to the predictions of the epidemic model. To explain these findings, I present studies examining the mechanisms of information diffusion, specifically how people respond to multiple exposures to information. The key finding of these studies---that central individuals in an online social network are less susceptible to becoming ``infected''---is sufficient to explain why social contagions fail to propagate. The studies also link the reduced susceptibility of central individuals to information overload and human reliance on cognitive heuristics to compensate for the brain's limited capacity to process information. %and its reliance on cognitive heuristics to deal with information overload.
Accounting for how people use cognitive heuristics to decide what information to pay attention to in social media dramatically simplifies dynamics of social contagion and allows for more accurate predictions of \remove{which memes}how far information will spread online.

\section{Size of Social Contagions}
%\section{Measuring Contagions}
Empirical studies of information spread in social media have failed to observe outbreaks as large as those predicted by the independent cascade model~\cite{Versteeg11icwsm,Goel12,Bakshy11}.
%Researchers measured how far information spreads on the Twitter and Digg follower graphs~\cite{Lerman10icwsm} and a variety of other online social networks~\cite{Goel12}.
This review focuses on two widely-studied social platforms---Twitter and Digg---although similar behaviors were observed in a variety of other social platforms~\cite{Goel12}. Twitter, a popular microblogging platform, allows registered users to broadcast short messages, or ``tweets.'' %, to followers.
These messages may contain URLs or descriptive labels, known as hashtags. In addition to composing original tweets, users can re-share, or ``retweet,'' messages posted by others.
%Although Digg had a different functionality and user interface, it had similar mechanics of information spread. %Digg's goal was to leverage its users' opinions to help people discover interesting news stories.
In contrast to Twitter, Digg focuses solely on news. Digg users submit URLs to news stories they find on the web and vote for, i.e., ``{digg},'' stories submitted by others. %, including the friends they followed.
Both platforms include a social networking component: users can subscribe to the feeds of other users to see the tweets those users posted (on Twitter) or the news stories they submitted or voted for (on Digg). The follow relationship is asymmetric; hence, we refer to the subscribing users as \emph{followers}, and the users they subscribe to as their \emph{friends} (or followees).

To measure the size of outbreaks on Twitter, researchers used URLs to external web content embedded in tweets as unique markers of information~\cite{Lerman10icwsm}. They tracked these URLs as users shared or retweeted the messages with their followers. %The number of people tweeting a URL who could be linked back to the source of information through follow links gave the size of the outbreak.
A similar strategy was used to track each news story on Digg.
Thus, the number of times a message containing a URL was retweeted or a news story was ``dugg'' in their respective networks gave an estimate of the outbreak size in that network.

\begin{figure}[t!]
\centering
\subfigure[Digg]{\label{subfig:digg}\includegraphics[width=0.4\textwidth]{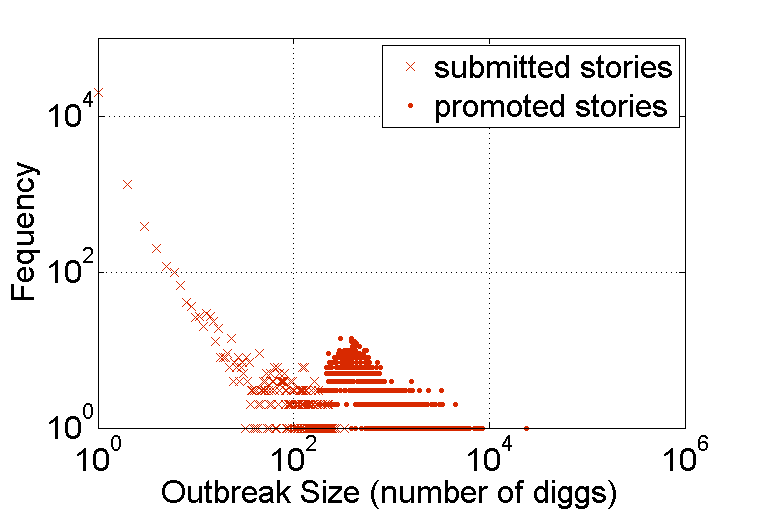}}
\subfigure[Twitter]{\label{subfig:twitter}\includegraphics[width=0.4\textwidth]{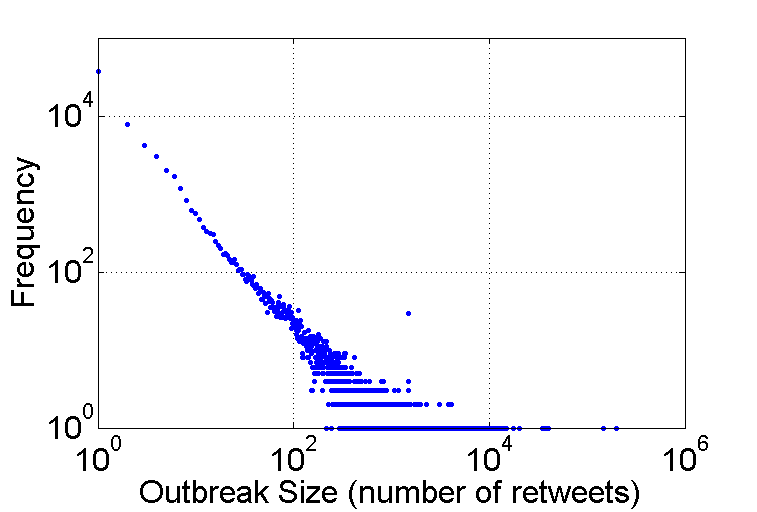}}
\caption{\emph{Size of outbreaks in social media.} Empirical measurements of the size of contagious outbreaks in social media sites (a) Digg and (b) Twitter. The x-axis represents the size of outbreak in terms of the number of people who dugg or retweeted specific information, and y-axis reports the frequency of events of that size.
}
\label{fig:outbreak}
\end{figure}

Figure~\ref{fig:outbreak} shows the distribution of outbreak sizes on the Twitter and Digg social platforms. Note that outbreaks have a long-tailed distribution, except for a bump on Digg that corresponds to promoted stories. When a newly submitted story accumulated enough votes, it was promoted to Digg's front page, where it was visible to everyone, not only to followers of voters~\cite{Hogg12epj}. The higher visibility of stories on the front page gave them a popularity boost, resulting in log-normally distributed popularity. However, even the most popular stories  did not penetrate very far. Only one story, about Michael Jackson's death, could be said to have reached ``viral'' proportions, i.e., reaching a non-negligible fraction of active Digg users (in this case, about 5\%). The next most popular story reached fewer than 2\% of Digg voters, and the vast majority of front page stories reached fewer than 0.1\% of the voters. Similarly,
%follower graph contained 700 thousand users with 36 million edges, yet
very few of the \remove{memes}outbreaks on Twitter reached more than 10,000 users, or less than 2\% of the active user population. These findings are in line with other studies, including by Goel et al.~\cite{Goel12}, who analyzed seven online social networks, ranging from communication platforms to networked games, to reach the same conclusion: the vast majority of outbreaks in online social networks are small and terminate within one step of the source of information.

\section{Mechanics of Contagion: Exposure Response}
\label{sec:exposure-response}
The observations above present a puzzle: what stops \remove{memes}information from spreading widely on social media? And why is outbreak size so much smaller than predicted by the independent cascade model?
A number of hypotheses could potentially explain the empirical findings:
\begin{description}
  \item[Subcriticality] %(\emph{ii})
  The vast majority of information spread is sub-critical, with transmissibility below the epidemic threshold. As a result, information is unlikely to spread upon exposure, and can be considered uninteresting. This hypothesis is easy to dismiss, since it is difficult to imagine that all the information shared on many different social media platforms is uninteresting.

  \item[Load balancing] %(\emph{ii})
  Social media users may modulate transmissibility of information to prevent too many \remove{memes}pieces of information from spreading and creating information overload. This hypothesis is difficult to evaluate, though it is not very credible, since such wide-scale coordination would be difficult to achieve. Moreover, it would require users to correctly estimate the popularity of \remove{memes}different pieces of information in their local neighborhood, a measurement that is easily skewed in networks~\cite{Lerman2016majority}.

  \item[Novelty decay] %(\emph{iii})
  Transmissibility of information could diminish over time as information loses novelty. A study~\cite{Hodas12socialcom} explicitly addressed this hypothesis, and found that the probability to retweet information on Twitter does not depend on its absolute age, but only the time it first appeared in a user's social feed.

  \item[Network structure] %(\emph{iv})
  Although it is conceivable that network structure (e.g., clustering or communities) could limit the spread of information, this hypothesis was ruled out~\cite{Versteeg11icwsm}. As can be seen in Figure~\ref{fig:icm}, the structure of the actual Digg follower graph somewhat reduces the size of outbreaks, but not nearly enough to explain empirical observations.

%\item[Individual heterogeneity]
%Central individuals with many friends are psychologically different from the general user population; in one or both of two possible ways:
%\begin{itemize}
%    \item they accumulate friends but ignore the content they post
%%  \item liking to share information about themselves with followers, but not interested in content from their followers: a lack of interest could be intrinsic to the person, not just due to overload of having too many followers. E.g., celebrity and fans
%
%  \item they like to read posts from many people but rarely post material themselves. (This seems unlikely -- how would users discover such people to follow in the first place -- and could be rejected based on observable behavior of central people.)
%\end{itemize}

  \item[Contagion mechanism]  %(\emph{v}) Finally,
  The decisions people make to vote for a story on Digg or retweet a URL on Twitter, once their friends have shared, it could differ substantially from the ICM. These differences could prevent information from spreading~\cite{Versteeg11icwsm}.

\end{description}

\begin{figure}[ht]
\begin{center}
\subfigure[Digg]{\label{subfig:digg-response}\includegraphics[width=0.4\textwidth]{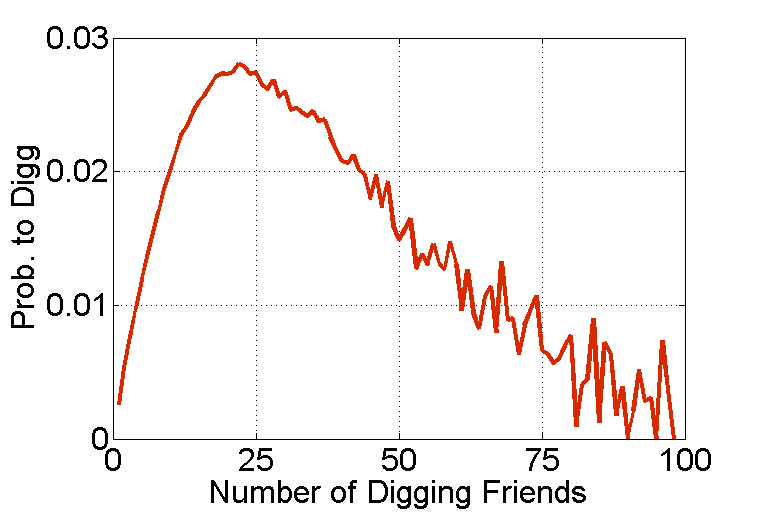}}
\subfigure[Twitter]{\label{subfig:twitter-response}\includegraphics[width=0.4\textwidth]{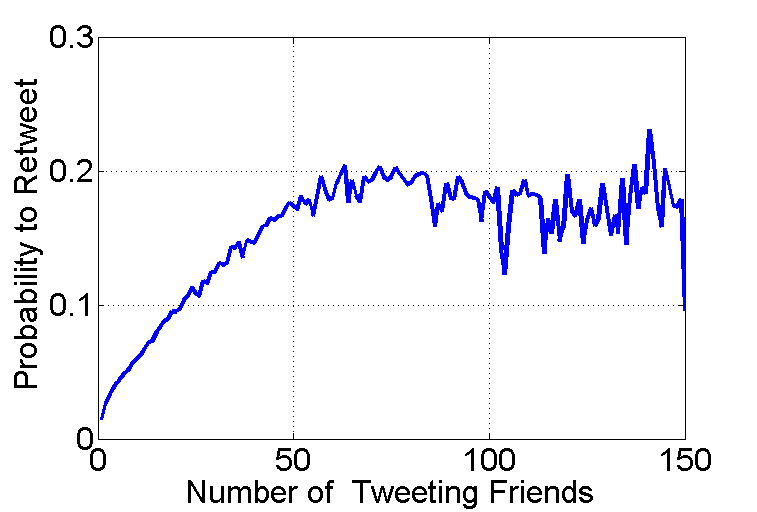}}
\end{center}
 \caption{\emph{The exposure response function for social media users.} The figures report the probability (averaged over all users) to respond to\remove{ a meme} information, i.e., (a) to digg a news story or (b) retweet a URL, as a function of number of friends who previously did so. \label{fig:exposureresponsefunctions} }
 \end{figure}

To characterize the mechanisms of contagion, researchers use the ``exposure response function.'' Since a person may be exposed to some information (or disease) by several friends, exposure response function gives the probability of an infection as a function of the number of exposures. Under the independent cascade model, infection probability rises monotonically with the number of infected friends as:
$
p_{\text{ICM}}(\text{infection}|k\ \text{exposures}) = 1 - (1 - \mu)^k,
$
where $\mu$ is the transmissibility. Using social media data, researchers empirically measured the exposure response function for Twitter and Digg users. To do this, they found all users who became ``infected'' (e.g., retweeted a URL~\cite{Hodas12socialcom} or adopted a hashtag~\cite{Romero11www} on Twitter, or ``dugg'' a story on Digg~\cite{Versteeg11icwsm}) after $k$ of their friends (i.e., the users he or she follows) became ``infected.'' The exposure response function is the ratio of the number of users who became ``infected'' to the number who did not become ``infected'' for different values of $k$.

Figure~\ref{fig:exposureresponsefunctions} shows the exposure response functions for Digg and Twitter, averaged over all users. The shape and magnitude of the exposure response functions are fundamentally different from that of the ICM. The form of the exposure response indicates that while initial exposures increase infection probability, additional exposures suppress new infections. According to Romero et al.~\cite{Romero11www}, such response is  suggestive of complex contagion, another popular model for describing social contagions, where ``infection'' does not occur until exposure by some specified fraction of friends~\cite{Granovetter78,Watts02,centola2007complex,centola2010spread}.
%Unlike complex contagion, larger fractions of infected friends appear to prevent new infections on these platform.

%
%\begin{figure}[ht]
%\begin{center}
%\includegraphics[width=0.4\textwidth]{figures/saturated_cascade_model}
%\end{center}
% \caption{Simulations of social contagion on the Digg follower graph while suppressing response to multiple exposures, as suggested by Fig.~\protect\ref{subfig:digg-response}.}\label{fig:fsm}
% \end{figure}
%

\begin{figure}[ht]
\begin{center}
\includegraphics[width=0.4\textwidth]{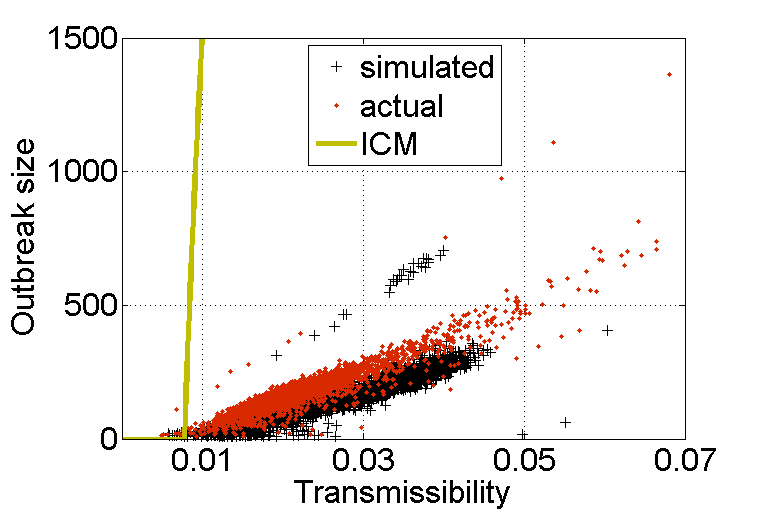}
\end{center}
 \caption{\emph{Size of simulated outbreaks on Digg as a function of transmissibility.}  Simulations of social contagion on the Digg follower graph (black crosses) using suppressed exposure response function suggested by Fig.~\protect\ref{subfig:digg-response}. Actual outbreaks on Digg are shown as red dots, while theoretically predicted (gold) line is the same as in Fig.~\protect\ref{fig:icm}. Suppressed response to repeated exposures vastly decreases the size of outbreaks as compared to prediction of the ICM (Fig.~\protect\ref{fig:icm}).
 \label{fig:fsm} }
 \end{figure}

Does the suppressed response to multiple exposures inhibit the spread of information online? Ver Steeg et al.~\cite{Versteeg11icwsm} simulated the spread of social contagions on the Digg follower graph with the suppressed exposure response function suggested by Fig.~\ref{subfig:digg-response}.
In the simulation, exposure response was approximated as follows.
If a node
%was exposed by an infected friend but
had any infected friends, it became infected with probability $\mu$. However, if it did not get infected (with probability  $1-\mu$), it was forever immune to new infections. Figure~\ref{fig:fsm} shows the size of the resulting outbreaks (red dots) as a function of transmissibility $\mu$.  The outbreaks are an order of magnitude smaller than those predicted by the independent cascade model, and in line with empirically observed outbreaks (black crosses).
This suggests that online contagions fail to spread due to the reduced susceptibility of social media users to multiple exposures.

\section{Limited Attention and Cognitive Heuristics}
\label{sec:heuristics}
Why don't social media users respond to multiple exposures to information?
In distinction to viral infection, social media users must actively seek out information and decide to share it before becoming ``infected.''
The enormous flux of information on social media  often saturates human ability to process information~\cite{Rodriguez14quantifying}.
%In addition, people often lack the time, motivation, or interest to process all the information in their social media feed.
Faced with an over-abundance of stimuli, humans evolved mechanisms to parsimoniously direct their attention to the most salient stimuli. What is salient depends on context: color, contrast, and motion help guide visual attention to important features of the environment, such as a predator. Social stimuli are also salient, as they aid coordination and help people avoid conflict. A variety of other cognitive heuristics are used to quickly (and unconsciously) focus attention on salient information~\cite{Kahneman73,Kahneman11}. In the context of social media, information that appears at the top of the web page or user's social feed is salient. As a result of this cognitive heuristic, known as ``position bias''~\cite{Payne51}, people pay more attention to items at the top of a list than those in lower positions. Social influence bias, communicated through social signals, helps direct attention to online content that has been liked, shared or approved by many others~\cite{Salganik06,Hogg2015hcomp}.

%Social influence, \emph{aka} ``bandwagon effect'', is one such heuristic: people pay attention to the choices of others. Therefore, displaying the number of people who have earlier shared some information is likely to affect how a persons reacts to it~\cite{Salganik06,Hogg2016hcomp}.  Another important heuristic in social media use is the ``position bias''~\cite{Payne51}: people pay more attention to items at the top of the list of items than those below. Hence, when a social feed is ordered chronologically, people will first view the most recent items near the top of the feed.
\begin{figure}[ht]
\begin{center}
\includegraphics[width=0.4\textwidth]{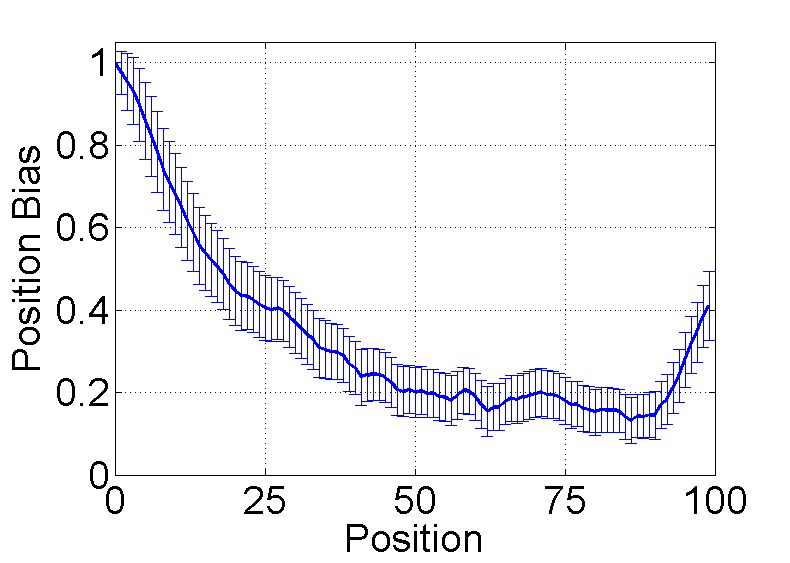}
\end{center}
 \caption{\emph{Position bias.} The relative decrease in recommendations received by items at different position within a list, compared to expected recommendations. Items in top positions (1--5) receive four to five times as much attention as items in lower positions (50--75). The rise near the end of the list is created by people who start inspecting the list from the end.	
 \label{fig:visibility} }
 \end{figure}

%perceptual saliency

Cognitive heuristics interact with how a web site displays information to users to alter the dynamics of social contagion. Twitter presents friends' messages in a user's feed as a chronologically ordered queue, with the most recently tweeted messages at the top. (Similarly, Digg orders news stories submitted by friends in a reverse chronological order.) Due to position bias, a user is more likely to see newest messages at the top of the feed than older messages in lower positions. Researchers conducted controlled experiments on Amazon Mechanical Turk to quantify position bias~\cite{Lerman14plosone}. They presented study participants with a list of 100 items and asked them to recommend those they found interesting. Figure~\ref{fig:visibility} shows the relative decrease in recommendations received at each list position, compared to what the items shown in those positions are expected to receive.
Items in top list positions (0--5) receive three to five times as much attention as those in lower positions (40--75), purely by virtue of being in those positions.

\begin{figure}[ht]
\begin{center}
\subfigure[Digg]{\label{subfig:digg-visibility}\includegraphics[width=0.4\textwidth]{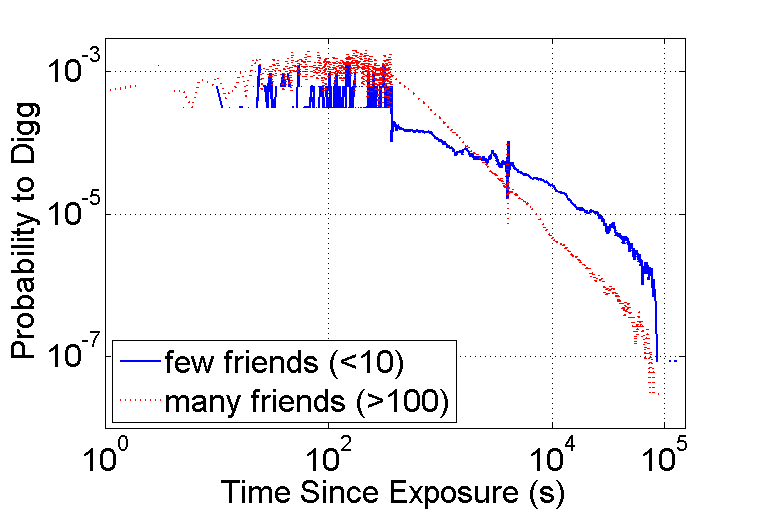}}
\subfigure[Twitter]{\label{subfig:twitter-visibility}\includegraphics[width=0.4\textwidth]{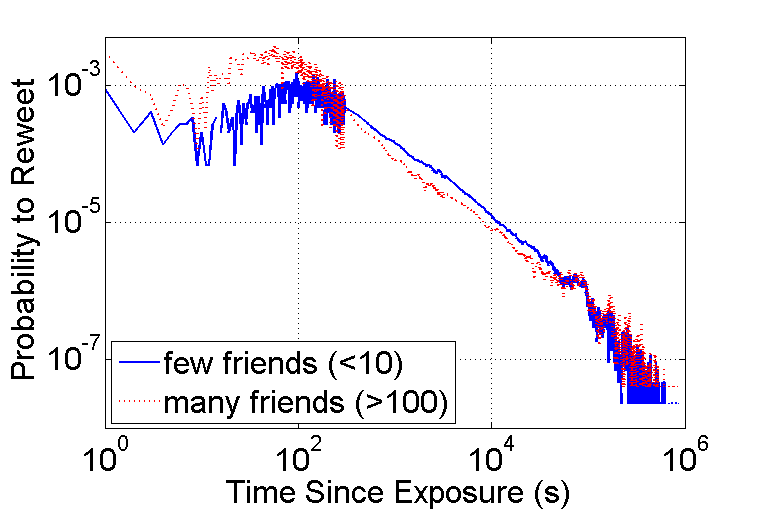}}
\end{center}
 \caption{\emph{Time response function for social media.} The probability to (a) digg a news story and (b) retweet a URL, as a function of the time since exposure, i.e., message's arrival in the user's feed. To remove the confounding effects of multiple exposures, the figure only considers single exposure items. Digg stories were only followed until promotion (the first 24 hours) during which time they were only visible to the followers. The data are smoothed using progressively wider smoothing windows, as in~\cite{Hodas12socialcom}.	
 \label{fig:time-response} }
 \end{figure}

However, in the observational data from social media, we do not know the feed position of a message at the time the user responds to it. Instead, we know that its position should be proportional to its age, i.e., time since its arrival in the user's feed, when the latter is a queue ordered by time of item's arrival. Figure~\ref{fig:time-response} confirms the effect. It shows the probability to digg (or retweet) an item as a function of the time since item's arrival. Though Twitter and Digg differ substantially in their functionality and user interface, they behave very similarly. The probability on both sites drops precipitously with time, which suggests that social media users are far less likely to see---and retweet---older messages in lower feed positions than newer messages in top feed positions.

% role of network structure/number of friends
Cognitive heuristics also interact with network structure of alter the dynamics of social contagion. The visibility of an item in a feed of a well-connected user (who follows many others) decreases faster in time than the visibility of an item in the feed of a poorly-connected user (who follows few friends).
As a result, well-connected users with more friends rarely retweet old content.
This is because these users receive many newer messages from their multitude of friends, which quickly push a given item further down the queue, where it is less likely to be seen. In contrast, poorly-connected users receive few messages, so that the visibility of an item does not decay as quickly. This effect is evident in Figure~\ref{fig:time-response}, where the probability to retweet (or digg) an item decreases faster for well-connected users (with more than 250 friends) than for the poorly-connected users (with fewer than 10 friends).

\begin{figure}[ht]
\begin{center}
%\subfigure[Twitter]{\label{subfig:twitter-response-breakdown}
\includegraphics[width=0.4\textwidth]{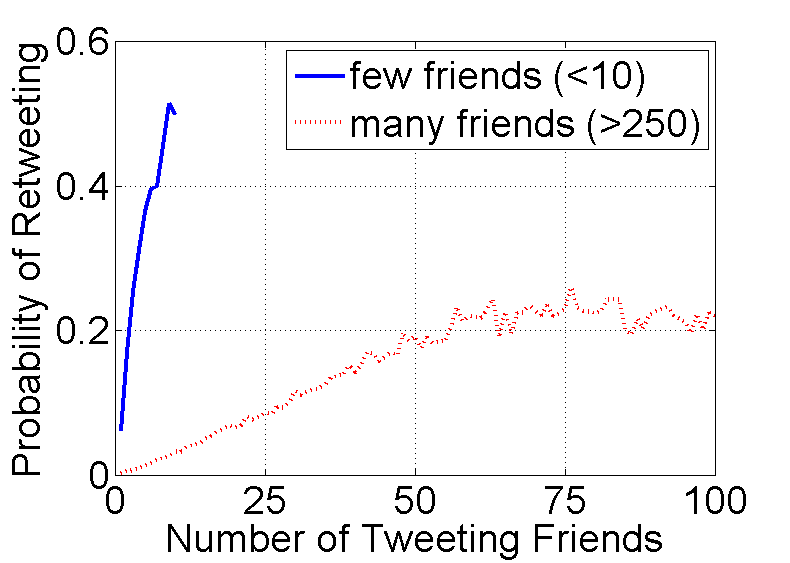}
%}
\end{center}
 \caption{\emph{Exposure response function for Twitter.} The figure shows the average probability to retweet\remove{ a meme} some message as a function of the number of friends who previously tweeted it for two classes of users, separated according to the number of friends they follow. The well-connected users with many friends are less likely to retweet; hence, they are less susceptible, than users with few friends. \label{fig:twitter-response} }
 \end{figure}

The quickly decaying visibility of information in the feeds of well-connected social media users reduces their susceptibility to becoming ``infected'' by that information\remove{those memes}. Figure~\ref{fig:twitter-response}  shows the exposure response function for two classes of Twitter users: those with few friends and those with many friends. The response probability of well-connected users with many friends is much lower compared to poorly-connected users, consistent with the argument that they have a harder time finding specific messages in their long feeds. Note also that both response functions increase monotonically, in line with simple contagion models, such as the ICM. The non-monotonic behavior observed for adoption of hashtags on Twitter~\cite{Romero11www} and in Figure~\ref{fig:exposureresponsefunctions} does not represent complex contaction, but is simply an artifact of averaging over heterogeneous populations of users with different cognitive load, i.e., different volumes of information in their feeds. Averaging the curves in Figure~\ref{fig:twitter-response}, we will observe an exposure response that initially increases, since both classes of users contribute; however, as the number of ``infected'' friends increases further, only users with many friends contribute to the response, bringing the average response function down.
This is an illustration of ``heterogeneity's ruses''~\cite{Vaupel85heterogeneity}: averaging over heterogeneous populations, each with its own behavior, can produce nonsensical behavioral patterns. When studying social systems, one needs to isolate the more homogeneous populations and carry out analysis within each population~\cite{Kooti16wsdm}.
%Only users with many friends contribute to portions of the response curve with large numbers of ``infected'' friends, bringing the average response function down.

\section{Predicting Social Contagions}
\label{sec:prediction}
% predicting retweets
Knowing how cognitive heuristics constrain user behavior enables us to more accurately predict social contagions.
To become ``infected,'' a user must first discover at least one message containing the information\remove{meme}.
We approximate a message's visibility using the  time response function, of the kind shown in Fig.~\ref{fig:time-response}, that gives the probability that a user with  $n_f$ friends  retweets or votes at a time $\Delta t$  after the exposure~\cite{Hodas12socialcom}.

To understand the dynamics of social contagion, we must also specify how users respond to multiple exposures to information. Here, the details of how the web site presents information matter. Twitter puts each newly retweeted item---the new exposure---at a top of the followers' feeds, creating a new opportunity for the followers to discover the item. Thus, if $k$ friends tweet some information, it will appear in a user's feed $k$ times in different positions. In contrast, Digg does not change the news story's relative position after a friend's digg, but increments the number of recommendations shown next to the story: after $k$ friends digg a story, it still appears only once in the user's feed, but with the number $k$ next to it. This number serves as a social signal that change a user's response. The effect of social signals on user's likelihood to become ``infected'' can be measured experimentally~\cite{Hogg2015hcomp} or estimated from observational data~\cite{Hodas14srep}.

\begin{figure}[ht]
\begin{center}
\subfigure[Digg]{\label{subfig:digg-prediction}\includegraphics[width=0.4\textwidth]{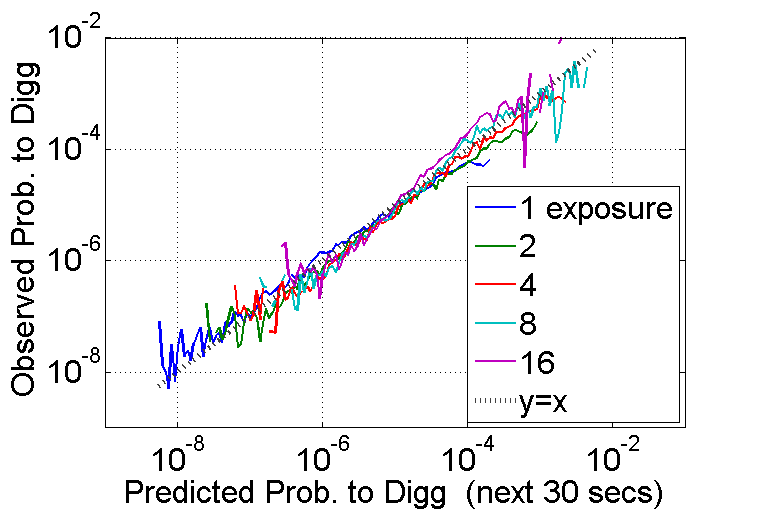}}
\subfigure[Twitter]{\label{subfig:twitter-prediction}\includegraphics[width=0.4\textwidth]{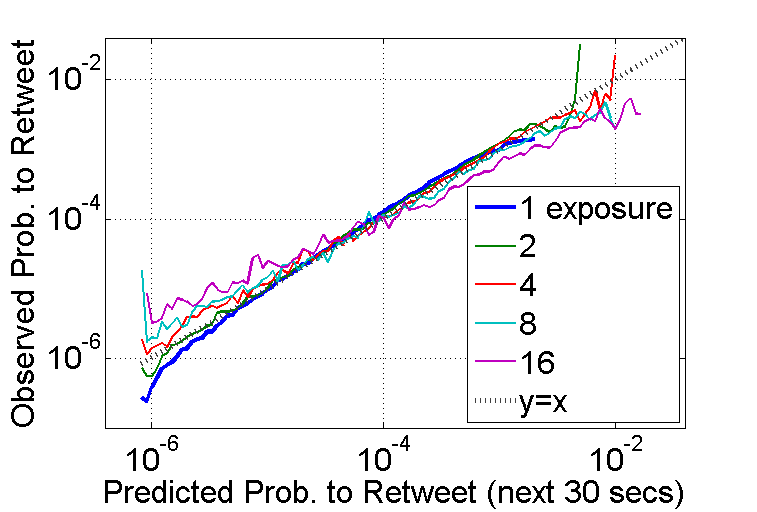}}
\end{center}
 \caption{\emph{Predicting response.} Observed probability to (a) digg or (b) retweet an item as a function of predicted probability, for different numbers of exposures to the item. \label{fig:prediction} }
 \end{figure}

Putting these factors together, \cite{Hodas14srep} proposed a simple model of social contagion where each exposure\remove{ to a meme} can independently cause an ``infection'' (i.e., a retweet). In contrast to the plain ICM, ``infection'' probability depends on the visibility of the exposures, which is related to the time of the exposures on Twitter or the time of the first exposure on Digg. A social signal, if present, will amplify ``infection'' probability.
To validate this model, the authors used it to forecast ``infections'' and compared them to observed ``infections.'' Specifically,  they calculated on a minute-by-minute basis the observed frequency that a user with some number of exposures in their feed retweeted\remove{ a meme} some specific information on Twitter  or dugg a story on Digg in the subsequent 30 seconds. Then they calculated the theoretical probability that the same user would act in those 30 seconds, given the same exposures. Figure~\ref{fig:prediction} shows the observed vs predicted probability of those infections, for different numbers of exposures in the users' feeds. For reference, perfect forecasts lie along the $y=x$ line. The unbiased fidelity of the proposed model suggests that once visibility of the exposures is taken into account, social contagion operates as a simple contagion, i.e., with infection probability increasing monotonically with the number of exposures. Other works incorporated visibility into models of user behavior that account for user interests~\cite{Kang15icwsm} and sentiment~\cite{Hogg13socialcom} about topics, their limited attention~\cite{Weng:2012dd}, and the multiple channels for finding \remove{memes}information, such as on Digg's front page~\cite{Hogg12epj}.

\section{Discussion}
The notion that networks amplify the flow of information has ignited the imagination of researchers and public alike. The few success stories---songs and videos that have spread in a chain reaction from person to person to reach millions---keep marketers searching for formulas for creating viral campaigns. Success, however, is rare. Empirical studies of the spread of information in online social networks revealed that information rarely spreads beyond the source.
The search for answers as to why information fails to spread in social media has uncovered the vital role of brain's cognitive limits in social media interactions.
%brain's (limited) capacity for processing information plays in mechanisms of information diffusion.% and underscored the differences between viruses and \remove{memes}information.
%Because people usually lack the capacity to process all information they receive, they prioritize it, paying more attention to the most recent items in the top of their social feed and ignoring the rest. As individuals become better connected, and their feed grows, they are able to process an ever smaller fraction of their feed, thereby decreasing their probability of becoming infected.

These cognitive limits are what differentiates the spread of information from the spread of a virus, and they must be accounted for in models of information diffusion. Specifically, in order to spread some information on social media, a person first has to discover it in his or her social feed. Discovery depends sensitively on how the web site arranges the feed, the flux of incoming information, and the effort the person is willing and able to expend on the discovery process.
Moreover, as people add more friends, the volume of information they receive may grow superlinearly due to the friendship paradox in social networks~\cite{feld91} and its generalizations~\cite{Hodas13icwsm,Kooti14icwsm}: a person's friends are more active and post more messages on average, then the person himself or herself does.
As a result, the volume of information may inevitably exceed an individual's cognitive capacity, creating conditions for information overload~\cite{Rodriguez14quantifying}. %In fact, information overload may be the inevitable outcome of the interplay between networks and cognitive limits.
%Thus, although the more central individuals in a network (i.e., who follow many others) also tend to be more active~\cite{Kooti14icwsm}, they cannot compensate for the higher information load, tipping them into the overload regime.
To deal with information overload, people rely on cognitive heuristics to focus only on salient information. In the context of social media, this means paying attention to the most recent messages at the top of their feed, and ignoring the rest. This reduces the probability that highly connected people will see and spread any given \remove{meme}piece of information in their feed, making them less susceptible to becoming ``infected.'' The reduced susceptibility of central users suppresses the spread of social contagions in social media. Accounting for these phenomena in models of information diffusion allows us to more accurately predict how far information will spread online.

The interplay between networks and human cognitive limits may have other non-trivial consequences. Potentially, people who have higher capacity to process information may put themselves in network positions allowing them greater access to information~\cite{Aral11,Kang15icwsm}, which they may then leverage for personal gain~\cite{Granovetter73,Burt95}. Understanding the role of social networks and cognitive heuristics and biases in individual and collective behavior remains an open research area.

\subsection*{Acknowledgments}
I am much indebted to my collaborators: Rumi Ghosh, Nathan O. Hodas, Tad Hogg, Jeonhyung Kang, Farshad Kooti, Laura M. Smith, Greg Ver Steeg, and Linhong Zhu. Their ability see farther and think deeper helped identify and unravel the puzzles that are the topic of this paper.
This work was funded, in part, but the Army Research Office under contract W911NF-15-1-0142 and by the National Science Foundation under grant SMA-1360058.

%\bibliographystyle{mdpi}
%\bibliography{../references,../../../lerman}

\end{document}